\documentclass[preprint]{aastex}
\usepackage{amssymb, amsmath}
\usepackage{arydshln}
\usepackage{color}
\usepackage[caption=false]{subfig}
\usepackage{rotating}
\usepackage[normalem]{ulem}

\newcommand{\cmnt}[1]{}

\newcommand{\acoupling}{a_{\Delta i \Delta j}}
\newcommand{\changetext}[1]{{\textcolor{black}{#1}}}
\newcommand{\suma}{{\Sigma_a}}

\begin{document}

\title{Brighter-fatter effect in near-infrared detectors -- II. Auto-correlation analysis of H4RG-10 flats}

\date{\today}
\author{Ami Choi}
\email{choi.1442@osu.edu}
\author{Christopher M. Hirata}
\affil{Center for Cosmology and AstroParticle Physics, The Ohio State University, 191 West Woodruff Avenue, Columbus, Ohio 43210, USA}

\begin{abstract}

The Wide Field Infrared Survey Telescope (WFIRST) will investigate the origins of cosmic acceleration using weak gravitational lensing at near infrared wavelengths. Lensing analyses place strict constraints on the precision of size and ellipticity measurements of the point spread function.  WFIRST will use \changetext{infrared} detector arrays, which must be fully characterized to inform data reduction and calibration procedures such that unbiased cosmological results can be achieved.  Hirata \& Choi 2019 introduces formalism to connect the cross-correlation signal of different flat field time samples to non-linear detector behaviors such as the brighter fatter effect (BFE) and non-linear inter-pixel capacitance (NL-IPC), and this paper applies that framework to a WFIRST development detector, SCA 18237.  We find a residual correlation signal after accounting for classical non-linearity.  This residual correlation contains a combination of the BFE and NL-IPC;  however, further tests suggest that the BFE is the dominant mechanism. If interpreted as a pure BFE, it suggests that the effective area of a pixel is increased by $(2.87\pm0.03)\times 10^{-7}$ (stat.) for every electron in the 4 nearest neighbors, with a rapid $\sim r^{-5.6\pm0.2}$ fall-off of the effect for more distant neighbors. We show that the IPC inferred from hot pixels contains the same large-scale spatial variations as the IPC inferred from auto-correlations, albeit with an overall offset of $\sim 0.06\%$. The NL-IPC inferred from hot pixels is too small to explain the cross-correlation measurement, further supporting the BFE hypothesis. This work presents the first evidence for the BFE in an H4RG-10 detector, demonstrates some of the useful insights that can be gleaned from flat field statistics, and represents a significant step towards calibration of WFIRST data.

\end{abstract}


\section{Introduction}
\label{sec:intro}

Weak gravitational lensing (WL) is one of the primary tools the Wide Field Infrared Survey Telescope (WFIRST) will use to detail the history of cosmic expansion and structure growth.  WL requires high fidelity measurements of galaxy shapes, which for WFIRST will be made on near infrared detector arrays consisting of a Teledyne H4RG-10 readout integrated circuit hybridized to $2.5$ $\mu$m cutoff HgCdTe.\footnote{See \citet{2007SPIE.6690E..0CL} and \citet{2014PASP..126..739R} for descriptions of this technology.}  The WFIRST Science Requirements Document specifies that the point spread function \changetext{(PSF)} ellipticity must be known with an error of $\le 0.057\%$ (RMS per component, in the convention of \citealt{2002AJ....123..583B}) and the size must be known with an error of $\le 0.072\%$ (trace of the 2nd moment matrix). As near infrared arrays have not hitherto been applied to a cosmological lensing analysis of this stringency, all potential biasing effects in the observational procedure must be thoroughly characterized.  Non-linear behaviors of particular interest include the so-called ``brighter fatter effect'' \cite[BFE;][]{2014JInst...9C3048A}, whereby a brighter source produces a larger image due to self-repulsion of electrons within a given pixel.  The BFE has recently been shown to affect the Dark Energy Survey, one of the current generation of ground-based optical lensing surveys \citep{2015JInst..10C5032G}.

These \changetext{infrared detector arrays} suffer from electrical cross-talk between the pixels, and this inter-pixel capacitance (IPC) dominates the flat field auto-correlation function \citep{2004SPIE.5167..204M}. Furthermore, the IPC can have a component that is signal-dependent (NL-IPC), as shown by \citet{Cheng2009, 2016SPIE.9915E..2ID, 2017OptEn..56b4103D, 2018PASP..130g4503D}.  \changetext{While IPC and NL-IPC have been investigated on a range of different infrared detector arrays,} most BFE studies in the literature focus on charge-coupled devices (CCDs), as most of the current and upcoming lensing surveys are or will be conducted in the optical regime \citep{2015ExA....39..207N,2015JInst..10C5024B, 2017JInst..12C3091L}.  \changetext{The most extensive previous efforts to characterize the BFE on infrared detectors were by} \citet{2017JInst..12C4009P,2018PASP..130f5004P}, who characterize the BFE on earlier generations of Teledyne detectors, H1RG and H2RG; the latter work uses a laboratory spot projector on a Euclid prototype H2RG, finding evidence consistent with a BFE.

\changetext{A complementary approach harnesses} the non-destructive read capability of the \changetext{HxRGs}, which is a powerful feature that enables multiple samples of a given flat or dark as a function of time.  Different frames can be cross-correlated, producing a signal that is sensitive to different types of detector effects.  Hirata and Choi (2019; hereafter Paper I) introduces a formalism for characterizing non-linear detector effects, including NL-IPC and the BFE in the cross-correlation signal of correlated double sample (CDS) images.  The aim of this work is to apply the methods of Paper I to laboratory data for a candidate WFIRST H4RG-10 sensor.

This paper is organized as follows. In \S\ref{sec:formalism}, we briefly summarize the theoretical predictions for the 2-point flat field correlation function contributions of IPC, BFE, and other detector effects from Paper I.  In \S\ref{sec:data}, we describe the data set for a H4RG-10 detector obtained from the Detector Characterization Laboratory at NASA Goddard Space Flight Center. In \S\ref{ss:char-basicadv} we characterize key properties such as gain, IPC, and non-linearity, and run tests to verify the robustness and repeatability of our analysis.  \changetext{In \S\ref{sec:bfe},} we present and compare \changetext{the main result of this paper:} measurements of the brighter-fatter effect.  We conclude and discuss areas of future exploration in \S\ref{sec:discussion}.

\section{Theoretical background}
\label{sec:formalism}

In this section, we briefly recall the formalism from Paper I used to describe the main detector effects of interest: the brighter-fatter effect, (non-linear) inter-pixel capacitance, and classical non-linearity.  We \changetext{describe} the main equations and parameters to orient the reader for the measurements and reference the details from Sections 2 and 3 of Paper I.  Table~\ref{tab:param_summary} provides a quick reference summarizing the detector parameters most relevant to this analysis. 

\begin{table}[]
    \centering
    \begin{tabular}{|c|c|l|}
    \hline\hline
    Quantity      &Units &Description  \\
    \hline
    Q   &ke   &Charge, current multiplied by time. \\
    \hline
   g &e/DN &Gain, corrected for IPC and classical non-linearity unless\\ & &specified (e.g. subscript `raw').     \\
   \hline
   K & &IPC kernel matrix, with $K_{0,0} = 1-4\alpha$, $K_{0,\pm 1} = K_{\pm 1,0} = \alpha$. \\
   \hline
   $\alpha$ &\% &Specifies the IPC kernel, average of horizontal (subscript `H')\\&& and vertical (subscript `V') components. Diagonal component\\ & &denoted with subscript `D'.\\
   \hline
  K$^{\prime}$ & &Signal level-dependent NL-IPC kernel matrix ($3\times3$).\\
  \hline
   $\beta$ &ppm/e & Leading order classical non-linearity coefficient. \\
   \hline
    $\acoupling$ & ppm/e &BFE kernel coefficients defined in terms of shifts from the \\&&central pixel ($\Delta i = \Delta j=0$). \\
    \hline
    $\suma$ & ppm/e &Sum of $\acoupling$ over $\Delta i$,$\Delta j$. \\
    \hline
    $[K^{2}a^{\prime}+KK^{\prime}]_{\Delta i, \Delta j}$ &ppm/e &Inter-pixel non-linearities (IPNL) including linear IPC, \\&&non-linear IPC, and BFE.\\
    \hline
    \end{tabular}
    \caption{Summary of detector parameters.}
    \label{tab:param_summary}
\end{table}

\subsection{Detector signals}

The observed signal $S$ in the detector is given in units of data numbers (DN), which are voltages quantized as 16-bit integers.  As the detector is exposed to light, the voltage across the photodiode decreases, causing $S$ to decrease.  In practice, the relation between the accumulated charge, $Q$, and the signal drop is non-linear and contains various contributions including IPC.  There is evidence that the IPC increases with increasing signal level, and this non-linear component, NL-IPC, is phenomenologically similar to the BFE in that a greater amount of coupling will also cause a larger change in FWHM in brighter stars.  However, the two effects occur in different stages of the signal measurement process (NL-IPC occurs in the conversion of charge to voltage, whereas the BFE occurs in the collection of charge) and imprint slightly different features on the flat field statistics, as we will investigate later on.

In the presence of IPC, the signal drop can be described at a pixel location $i,j$ (column and row indices) by:
\begin{equation}
S_{\rm initial}(i,j) - S_{\rm final}(i,j) = \frac1g \sum_{\Delta i,\Delta j} [K_{\Delta i,\Delta j} + K'_{\Delta i,\Delta j} \bar Q] Q_{i-\Delta i, j-\Delta j},
\label{eq:NL-IPC-general}
\end{equation}
where $\bar Q$ is the mean accumulated charge ($It$ in a flat exposure, with current $I$ per pixel given in units of e/s and time $t$ in seconds) and $g$ is the gain (units: e/DN).  ``Initial'' is defined here as $t=0$, or immediately following a reset. The kernel matrix ${\bf K}$, describing the IPC, satisfies the normalization $\sum_{\Delta i,\Delta j} K_{\Delta i,\Delta j} = 1$.
In the case where the cross-talk is equally distributed to the four nearest neighbors $K_{0,0} = 1-4\alpha$, $K_{0,\pm 1} = K_{\pm 1,0} = \alpha$, and all others are zero. However, asymmetries between the horizontal and vertical directions ($K_{0,\pm 1}\neq K_{\pm 1,0}$) are commonly observed, so we separately measure $\alpha_{\rm H} = K_{\pm 1,0}$ and $\alpha_{\rm V} = K_{0,\pm 1}$; if these are different then we define $\alpha$ to be their average $(\alpha_{\rm H}+\alpha_{\rm V})/2$. We also allow for diagonal IPC, $\alpha_{\rm D} = K_{\pm 1, \pm 1}$. In Equation~(\ref{eq:NL-IPC-general}), we parameterize the NL-IPC to be dependent on the mean signal level as the kernel matrix ${\bf K'}$.

We also allow for a classical (total count-dependent) non-linearity in the detectors. This is modeled by the mapping of charge $Q\rightarrow Q-\beta Q^2$, where $\beta$ (units: ppm/e) is the leading-order non-linearity coefficient. We perform this mapping {\em after} the IPC convolution. In reality non-linearity and IPC are happening at the same time, but in the case of small fluctuations around a mean signal (as occurs for a flat field) the ordering does not matter.

\subsection{The brighter fatter effect}

We use the \citet{2014JInst...9C3048A} model, in which the pixel areas are modified by existing charge in accordance with a kernel $\acoupling$:
\begin{equation}
{\cal A}_{i,j} = {\cal A}_{i,j}^0 \left[ 1 + \sum_{\Delta i,\Delta j} a_{\Delta i,\Delta j} Q({i+\Delta i,j+\Delta j}) \right],
\label{eq:Aij}
\end{equation}
where ${\cal A}_{i,j}$ is the effective pixel area, and ${\cal A}_{i,j}^0$ is the original pixel area.  We will quote $\acoupling$ in units of 10$^{-6}$ e$^{-1}$, ppm/e, or equivalently \%/$10^4$ e.  Note that while $\acoupling$ is formally dimensionless, the aforementioned choice of units is convenient because a measured value of $\acoupling$ in ppm/e maps into the expected order of magnitude of the effect on a star in percent.  In this work, we do not study the pixel-dependence of the BFE and also assume discrete translation invariance.  We also define $\suma = \sum_{\Delta i,\Delta j} a_{\Delta i,\Delta j}$; for a pure BFE (no signal-dependent QE) we should have $\suma=0$. As some of the tests we conduct later are not sensitive to $\suma$, we also define:
\begin{equation}
a'_{\Delta i,\Delta j} \equiv a_{\Delta i,\Delta j} - \delta_{\Delta i,0}\delta_{\Delta j,0} \suma.
\end{equation}
By construction, the $a'$ coefficients sum to zero.

\subsection{Correlation functions}

\changetext{As these infrared} detectors allow multiple samples up the ramp, correlations can be measured not only between different pixels but also between different frames.  The temporal structure of the correlations is key to disentangling the BFE and NL-IPC. The correlation function $C_{abcd}(\Delta i,\Delta j)$ responds to the BFE and NL-IPC as given by Eq. (\changetext{51}) in Paper I. $abcd$ are indices representing frames, and we assume that $a<b$ and $c<d$, since these functions contain all the information because of symmetries, but we do not assume anything else about the ordering. The exposure intervals $a...b$ and $c...d$ may be the same, may overlap, or may be disjoint.  For the purposes of Paper II, terms of order $\alpha$, $\alpha^2$, $\beta$, ${a}$, $\alpha\beta$, and $\alpha a$ are kept, while higher order terms are dropped.

First, we consider the non-overlapping correlation function, where $a<b<c<d$, which leverages the non-destructive read capability of the \changetext{infrared} detectors to determine whether there is a correlation between current fluctuations in an earlier part of an exposure and the current fluctuations in an adjacent pixel at a later part of the exposure and an anti-correlation in the same pixel.  This is Eq. (\changetext{58}) in Paper I, through which we can use the `observables' $C_{abcd}(\Delta i,\Delta j)$, $I$, $g$, $\alpha$, $\alpha_{\rm H}$, $\alpha_{\rm V}$, and $\beta$ to solve for the inter-pixel non-linear (IPNL) effects $[K^2a]_{-\Delta i,-\Delta j}+[KK']_{\Delta i,\Delta j}$.

\changetext{The following two special cases of interest can help distinguish between BFE and NL-IPC contributions to the IPNL effects measured by the non-overlapping correlation function. The first of these} is the equal-interval correlation function, where $a=c<b=d$ -- this is the auto-correlation of a single difference image $S_a-S_b$, which is most similar to the auto-correlation that one would obtain from a CCD.  The relevant equations in Paper I are given by Eqs (\changetext{52-57}).

The final case of interest involves $a=c<b<d$ in the mean-variance plot, which is a common diagnostic of the gain of a detector system.  The raw gain is more generally written as
\begin{equation}
\hat g^{\rm raw}_{abcd} \equiv \frac{M_{cd}-M_{ab}}{V_{cd}-V_{ab}},
\label{eq:g-raw}
\end{equation}
where $M_{ab} = \langle S_a(i,j)-S_b(i,j)\rangle$ and $V_{ab} = C_{abab}(0,0)$ is the variance of a difference frame. In Paper II we will consider only $a=c<b<d$, for which the expression for raw gain can be written as Eq. (\changetext{62}) from Paper I.
In this equation, there are two time-dependent terms within the curly brackets containing non-linear correction terms; the first involves the start time $t_a$, while the second depends on the duration time $t_{ad}+t_{ab}$.  In \S\ref{ss:mean-var-relation}, we revisit these time dependencies\changetext{, showing how they can inform our understanding of how the BFE and NL-IPC contribute to the measured }.

\section{Data}
\label{sec:data}

Dark and flat illumination frames were acquired for an H4RG-10 detector array labelled as SCA 18237 -- a 2.5 $\mu$m cutoff device with 10 $\mu$m pitch pixels -- at the Detector Characterization Laboratory (DCL) at NASA Goddard Space Flight Center. SCA 18237 was one of the arrays built for the WFIRST infrared detector technology milestone \#4 (yield demonstration). It also underwent environmental testing (technology milestone \#5: thermal cycling and vibration) and showed no performance degradation.\footnote{The technology milestone reports are available at:\\ {\tt https://wfirst.gsfc.nasa.gov/science/sdt\_public/wps/references/WFIRST\_DTAC4\_160922.pdf} and\\ {\tt https://wfirst.gsfc.nasa.gov/science/sdt\_public/wps/references/WFIRST\_DTAC5\_nobackup.pdf}.}

There are some key differences between the detector operation in these tests and the planned operation in flight. Most notably, the data here were acquired with a laboratory controller (Gen-III Leach), rather than the ACADIA flight controller \citep{2018SPIE10709E..0TL}. Furthermore, the data were acquired in 64 output channel mode, whereas 32 output channels are planned for flight. Finally, the H4RG-10 has a guide window mode, which was not active during these tests but is planned for flight. Other data have been taken to assess the impact of the guide window on science performance, but are not presented in this study.

\begin{figure}
    \centering
    \includegraphics[width=6.4in]{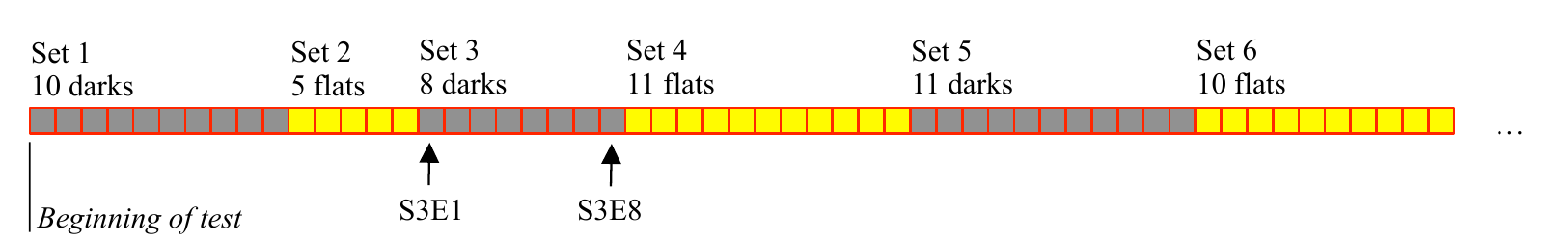}
    \caption{\label{fig:sequence}The sequence of exposures used in this test, containing interspersed darks and flats.}
\end{figure}

The data are provided in binary FITS format, with multiple up-the-ramp samples saved in 66 frames (total exposure time $\sim 182$ s) at the native 16 bit precision of the analog-to-digital converter (ADC) for a file size of 2.2 GB each. The dark and flat exposures were grouped into ``sets,'' with each set consisting of a sequence of back-to-back identical exposures, as shown in Figure~\ref{fig:sequence}.\changetext{\footnote{The sequence was originally designed to study persistence issues.}} The odd-numbered sets contained dark exposures, while the even-numbered sets contained flat field exposures, but the number of exposures in each set was varied to provide information on persistence and hysteresis. Each set has an exposure number, thus we refer to ``Set 1, Exposure 1'' (S1E1), S1E2, etc. We will often discuss the ``first flats'' in a sequence, indicating the first flat exposure following a set of darks: S2E1, S4E1, etc. In the presence of persistence and hysteresis effects, the first flats show a slightly different signal level and non-linearity curve than the subsequent flats. The ordering of the ADC levels was opposite from the formalism of this paper (i.e., the signal in DN increases during illumination) so we inverted the ordering, $S\rightarrow 2^{16}-1-S$, before any processing.\footnote{We have received data samples from the DCL in both increasing and decreasing formats, and so we implemented an inversion option in our routines to read the FITS files.}

The wavelength of illumination was 1.2 $\mu$m, and thus we do not expect to observe quantum yield effects.

\begin{figure}
    \centering
    \includegraphics[width=6.5in]{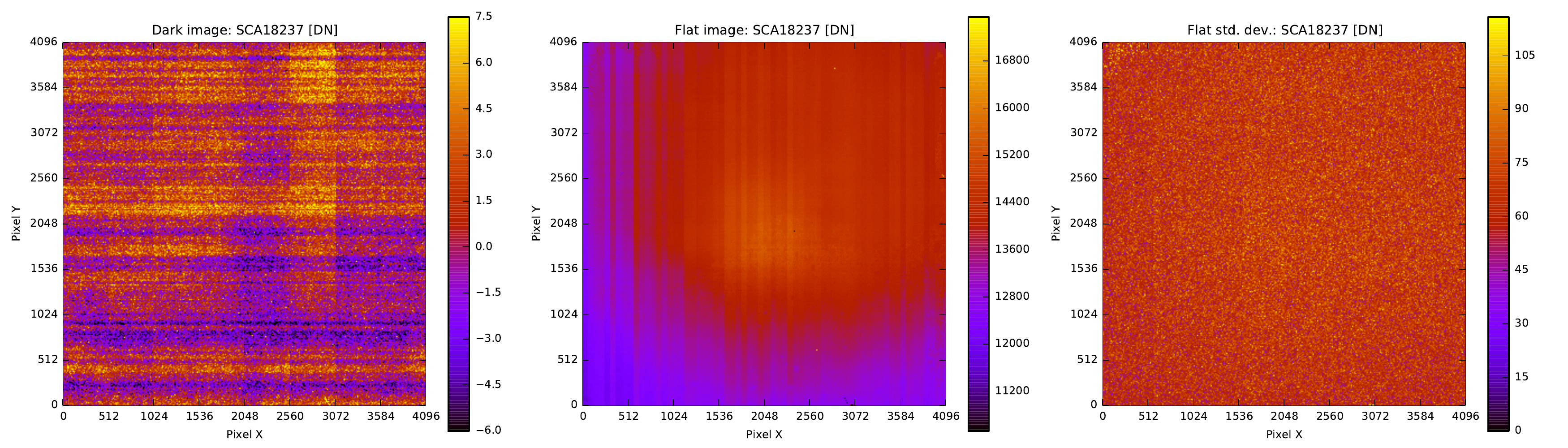}
    \caption{\label{fig:darkflat}Dark (left) and flat (middle) images from SCA 18237. The panels show the CDS images $S_1-S_{21}$, i.e., difference between 1st and 21st frame, and have been median-binned $4\times4$ on the spatial axes. The exposures used were S1E1 for the dark (left) and S2E1 for the flat (middle). These are raw images, in DN, and without reference pixel subtraction. The right panel shows the standard deviation of the difference image (S2E1$-$S4E1)$/\sqrt2$.}
\end{figure}

Some representative dark and flat images from the SCA are shown in Figure~\ref{fig:darkflat}. The left panel is a dark image (the CDS image $S_1-S_{21}$ for exposure S1E1), with the median taken in $4\times 4$ bins. Note the strong horizontal banding in the raw dark image, which motivates our choice of the reference pixels on the left and right sides of the arrays for the analyses in this paper. The middle panel shows a flat field image (CDS image $S_1-S_{21}$ for exposure S1E1, also $4\times 4$ median-binned). Some cosmetic defects can be seen. The right panel shows the flat field standard deviation. The CDS images $S_1-S_{21}$ were computed for two flats -- S2E1 and S4E1 -- and the normalized difference was taken, (S2E1$-$S4E1)/$\sqrt2$. In each $4\times 4$ bin, we computed the standard deviation of the 16 pixels. Some of the cosmetic defects are also visible in this image.

\section{Characterization based on flat fields}
\label{ss:char-basicadv}

\begin{figure}
    \centering
    \includegraphics[height=8in]{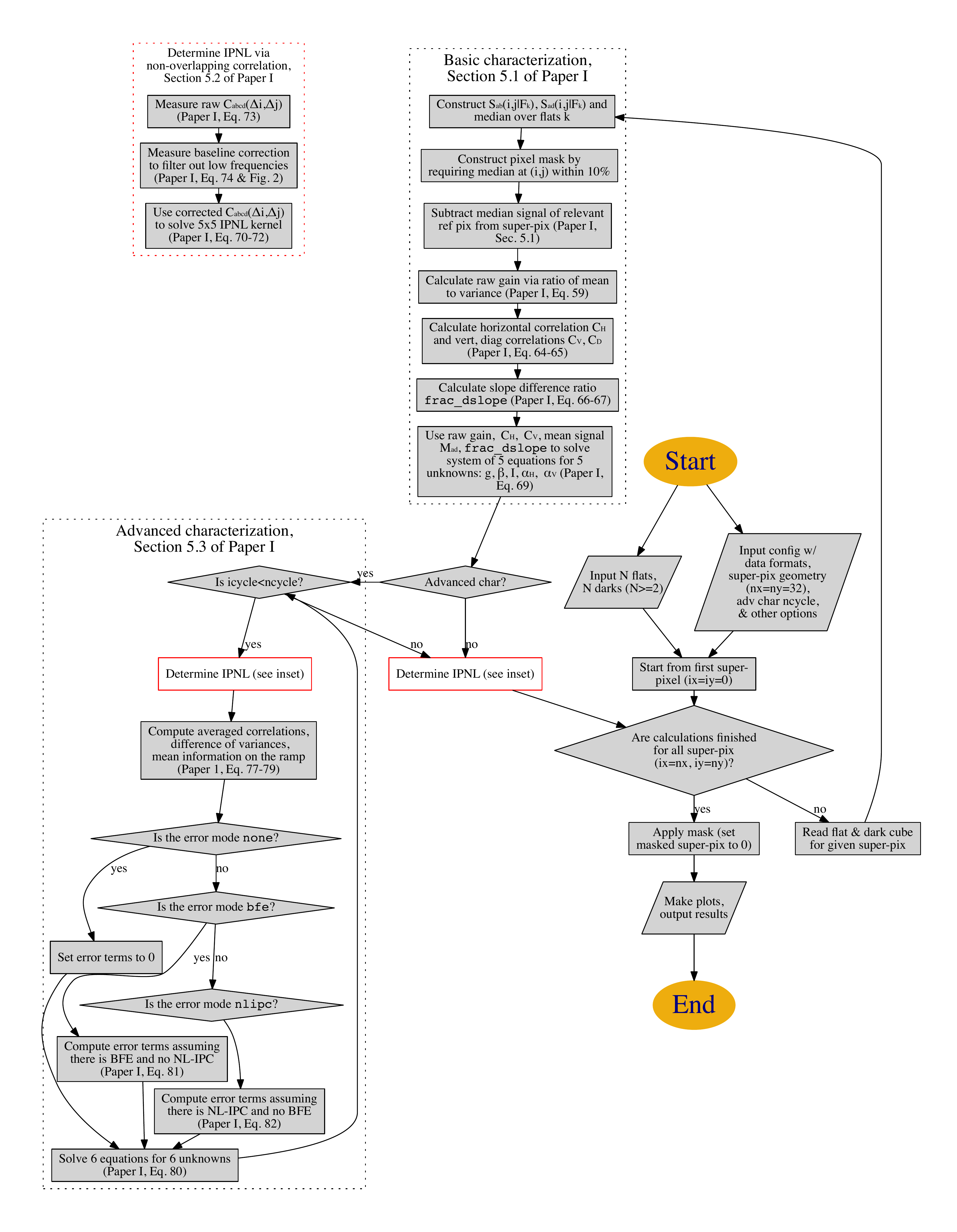}
    \caption{Flowchart showing overview of analysis procedure including basic characterization, IPNL calculation via the non-overlapping correlation function, and advanced characterization.  Full details can be found in the referenced equations and text of Paper I.}
    \label{fig:flow}
\end{figure}

As discussed in Section 5 of Paper I, we want to extract the calibration parameters ($g$, $\alpha$, $\beta$, $a_{\Delta i,\Delta j}$, etc.) from a suite of flat field and dark exposures for SCA 18237. We use \textsc{solid-waffle}, which is described in Paper I.  \changetext{We summarize the procedure in Figure~\ref{fig:flow}}.

\changetext{Figure~\ref{fig:flow} begins with input of} $N$ flat fields and $N$ dark images, where $N\ge 2$. The SCA is broken into a grid of $N_x\times N_y$ ``super-pixels,'' each of size $\Delta_x \times \Delta_y$ physical pixels. Statistical properties such as medians, variances and correlation functions are computed in each super-pixel. Note that $N_x\Delta_x = N_y\Delta_y = 4096$ for an H4RG (and 2048 for an H2RG). Super-pixels may be made larger to improve S/N, but this implies more averaging over the SCA so localized features and patterns may be washed out. Our default analyses have $N_x=N_y=32$, so there are $32^2=1024$ super-pixels, each containing $128\times 128$ physical pixels.

\changetext{Each super-pixel is passed through three main steps (grouped with dashed lines in Figure~\ref{fig:flow}): first, ``basic'' characterization, which measures the gain from the mean-variance plot and corrects it for the IPC (inferred from the CDS autocorrelation function) and the non-linearity $\beta$ (measured from curvature of the ramp); second, IPNL determination using the non-overlapping correlation function, i.e., $C_{abcd}(\Delta i,\Delta j)$ for $a<b<c<d$; third, advanced characterization, which iteratively removes the biases in gain, IPC, and non-linearity measurements caused by IPNL. We use the ``{\tt bfe}'' correction scheme for the advanced characterization, since our results show that the BFE dominates over NL-IPC as the main form of IPNL.}

\begin{figure}[t!]
\includegraphics[width=6.2in]{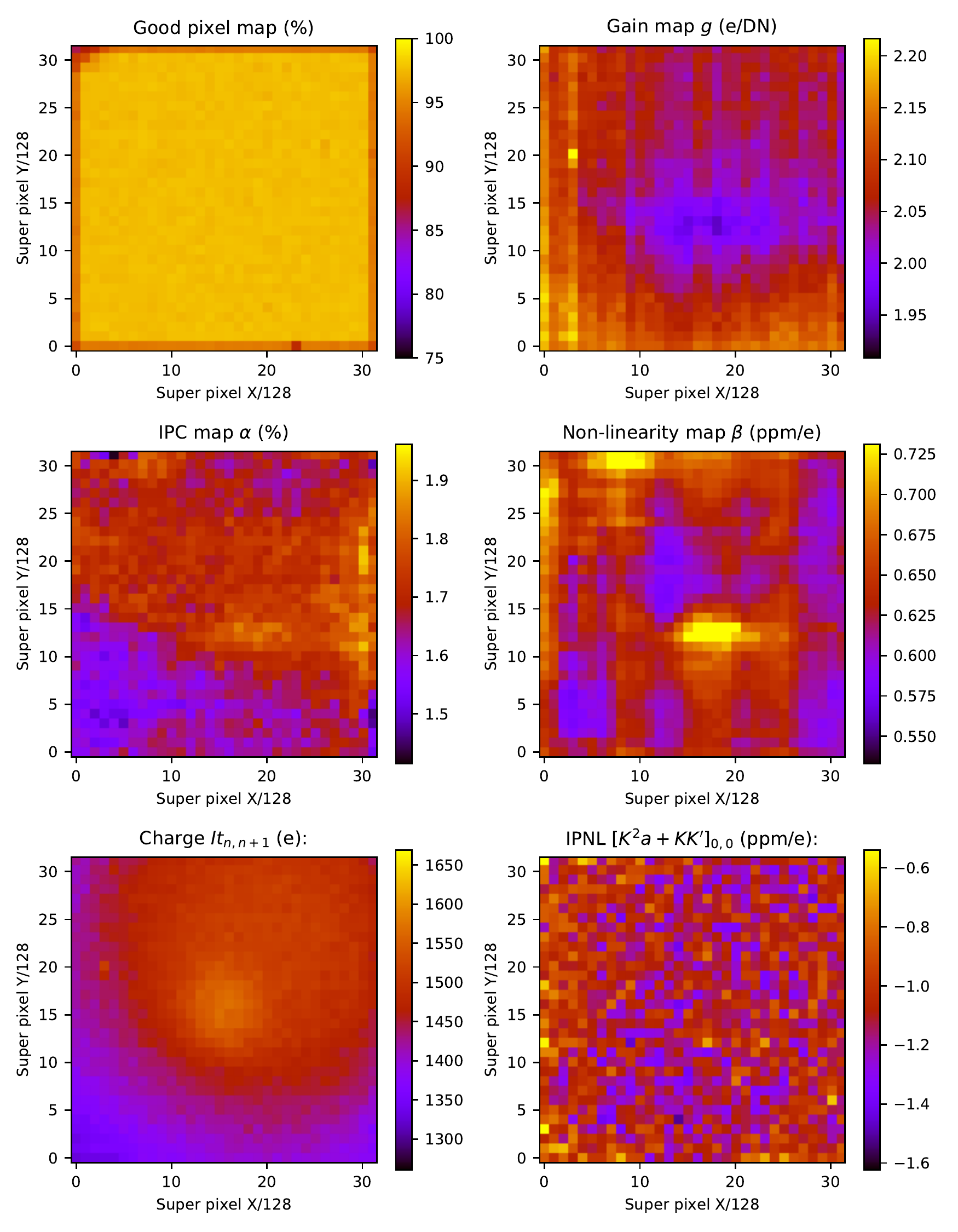}
\caption{\label{fig:ac18237} Advanced characterization for SCA 18237.}
\end{figure}

Figure~\ref{fig:ac18237} shows advanced characterization \changetext{results} for SCA 18237 (23 flats and darks). The figure shows good pixel percentages, $g$, $\alpha$, $\beta$, charge per time step, and the central kernel value of the inter-pixel non-linearities in each of the 1024 super-pixels.  We note that some spatial variation appears in the maps of $g$, $\alpha$, and $\beta$.  Additionally, the IPNL appears to be dominated by noise rather than real fluctuations across super-pixels.  The theoretical Poisson noise error on the IPNL for SCA 18237 is approximately equal to
\begin{eqnarray}
\sigma([K^2a'+KK']_{\Delta i,\Delta j}) &=&
\frac{1}{\sqrt{N_{\rm pix}(N_{\rm flat}-1)(It_{ab})(It_{cd})}}
= \frac{1}{\sqrt{128^2(23-1)(1463\times 8)^2}} 
\nonumber \\
&=& 1.42\times 10^{-7} = 0.142 {\rm\,ppm/e}, 
\end{eqnarray}
where $N_{\rm pix}$ is the number of pixels averaged together and $N_{\rm flat}$ is the number of flat fields used. The range of IPNL values in Figure~\ref{fig:ac18237} encompasses about 7$\sigma$.  We also verify that the measured standard deviation is 0.145 ppm/e, which is consistent with the predicted error.

\begin{table}
    \centering
    \begin{tabular}{ccccccc}
    \hline\hline
     Quantity     &Units  &\multicolumn{5}{c}{ncycle} \\
      & & 0 &1 &2 &3 &4\\
      \hline
     Gain $g$               &e/DN  &2.1574  &2.0636  &2.0643 &2.0643 &2.0643\\
     & &0.0022 &0.0016  &0.0016  &0.0016 &0.0016\\

     IPC $\alpha_{\rm H}$&\%  &2.0761 &1.6301  &1.6330 &1.6330 &1.6330 \\
     & &0.0138 &0.0070  &0.0046 &0.0046 &0.0046\\
     IPC $\alpha_{\rm V}$&\%  &2.2023 &1.7410  &1.7439 &1.7439 &1.7439  \\
     & &0.0100 &0.0053 &0.0053  &0.0053 &0.0053 \\
     IPC $\alpha_{\rm D}$&\%  &- &0.1881  &0.1881 &0.1881 &0.1881   \\
     & &- &0.0029  &0.0029  &0.0029 &0.0029\\
     Non-linearity $\beta_{\rm ramp}$  &ppm/e  &0.5597 &0.5832  &0.5830 &0.5830 &0.5830   \\
     & &0.0010 &0.0010  &0.0010  &0.0010 &0.0010 \\
     \hline
            \end{tabular}
    \caption{Statistics of gain, IPC, and ramp curvature as a function of \texttt{ncycle} iteration for a stack of 3 SCA 18237 flats. Two rows of values are provided for each quantity with the first row corresponding to the mean value over all good super-pixels and the second row corresponding to the standard error on the mean.  The measurements converge by three iterations.}
    \label{tab:ncycle}
\end{table}

Table~\ref{tab:ncycle} shows the difference in the main quantities of interest for SCA 18237 as a function of the number of iterations of correction for IPNL.  For each quantity, the first row corresponds to means over $N_{good}$ good super-pixels, and the second row gives statistical uncertainties computed as standard deviations on the mean of $N_{good}$.  The post-$\alpha\beta$-corrected gain variances are also smaller than the variances of the raw gain.  Choosing \texttt{ncycle=3} and above yields the same values as presented in Table~\ref{tab:ncycle} for \texttt{ncycle=2}.  For the remainder of this work, we will use advanced characterization with \texttt{ncycle=3}.

\subsection{Robustness against biases}
\label{sec:verify}

We have run the characterization for a number of configurations designed to check the stability and reproduceability of the $g$, $\alpha$, and $\beta$ parameters.  The means and standard deviations of these values are given in Table~\ref{tab:sca18237_full}.  We also provide a systematic uncertainty for $\beta_{ramp}$ to account for its stability.  We compute this as the sample standard deviation of all of the good super-pixel measurements.

For every measurement in Table~\ref{tab:sca18237_full}, we show three results. The first is ``1st,n3,'' which is based exclusively on the first flat illumination in a set (3 flats: S2E1, S4E1, S6E1). These should be the least affected by persistence/hysteresis effects. The second result is ``2nd,n3,'' which is based on the second flat illumination (3 flats: S2E2, S4E2, S6E2). This will be more strongly affected by persistence and hysteresis from the previous illumination, but less affected by stability issues associated with the flat lamp turning on. The third result is ``fid,n23,'' which contains 23 flat fields (S2E[1-5], S4E[4-11], S6E[1-10])\changetext{, the subset of the flat field data set to which we had access at the time of the analysis.\footnote{One may notice that Figure~\ref{fig:sequence} shows a total of 26 flats, and that S4E[1-3] were not included in the ``fid,n23`` case -- this was due to data transfer issues that were only resolved later.}} This mixes first and subsequent flats, but has the greatest statistical power. We see that the non-linearity $\beta_{\rm ramp}$ changes substantially depending on this test, with the ``1st,n3'' case giving a result 0.05 ppm/e lower than subsequent flats (i.e., the first ramp is more linear than the second ramp); moreover the inferred charge in the first ramp is $0.55\pm0.15$\%\ lower. We are continuing to study how much of this is due to the detector and how much to the test setup. However there is no detectable change in the gain, IPC, or IPNL coefficients in the 1st vs.\ subsequent flats.

We next vary the quantile level used for estimating the variance in gain. The default is 75 (inter-quartile range: $\pm 0.67\sigma$ for a Gaussian), and we compare setting this to 85 (i.e., estimating the variance from the difference between 15th and 85th percentiles: $\pm1.04\sigma$ for a Gaussian). We would expect the gains to change if the variance measurement is biased by non-Gaussianity of the signal (e.g., kurtosis or outliers). The gains change by $<0.03$\%.

Next, we consider the clipping fraction $\epsilon$ for the IPC correlations.  The default is 0.01, which clips the top 1\% and bottom 1\% of the pixels before computing a covariance.  We set $\epsilon=0.025$ to check the impact of clipping the top and bottom 2.5\%. This is a consistency check for the clipping correction factor in Appendix A of Paper I, which changes from $f_{\rm corr}=0.7629$ to $f_{\rm corr}=0.5758$. Despite this large change ($|\Delta f_{\rm corr}|/f_{\rm corr}=0.25$), the change in IPC is only $|\Delta\alpha|/\alpha \approx 0.004$.

We then investigate turning full reference pixel corrections on and off. $\beta$ becomes slightly smaller for all three sets of flats, ranging from 1.6\% to 2.8\% compared to the fiducial setup.

We also check that our choice of the $\texttt{bfe}$ error mode in the advanced characterization scheme does not affect the output $\beta$ and BFE+NL-IPC coefficient values by running the same analysis using the $\texttt{none}$ and $\texttt{nlipc}$ error modes.  The resulting values are shown in Table~\ref{tab:sca18237_full} and do not show strong deviations from the fiducial setup that uses the  $\texttt{bfe}$ error mode.

Finally, the default calculation uses the first frame as the reference ($t=0$). The first frame in the data cube is, however, 1 frame after the reset frame, so 2.75 s after the reset. This means the gain computed, in e/DN, is in fact {\em not} the slope of the charge vs.\ signal curve at the reset level, but the charge one frame later. We did one run where the ``reference'' ($t=0$) is set to frame 0 (the reset frame) instead of frame 1. We expect most parameters such as the IPC and IPNL to not change, but we do expect the gain to change in accordance with
\begin{equation}
\Delta g = g|_{\rm frame\,0} - g|_{\rm frame\,1} \approx -2\beta g It_{0,1}.
\label{eq:DG}
\end{equation}
We find that with this change, the changes in IPC are $\Delta\alpha=0.0005\%$; in IPNL are $\Delta [K^2a'+KK']_{0,0}=0.006$ ppm/e; and in $\beta_{\rm ramp}$ are $\Delta\beta_{\rm ramp}=-0.001$ ppm/e. The expected changes in gain do occur: they are $\Delta g = -0.0036$ e/DN (measured), versus $-0.0035$ e/DN (expected from Eq.~\ref{eq:DG}).

\begin{table}[]
\scriptsize  
    \centering
    \begin{tabular}{llccccccc}
    \hline\hline
     Quantity     &Units  &\multicolumn{3}{c}{Flat type, number} &\multicolumn{3}{c}{Uncert} &Notes \\
      & & 1st,n3 &2nd,n3 &fid,n23 &stat.(3) &stat.(23) &sys.(3) & \\
      \hline
     Charge, $It_{n,n+1}$    &ke   &1.4536  &1.4617  &1.4632  &0.0016  &0.0015 &\\
     Gain $g$               &e/DN  &2.0643  &2.0652  &2.0639  &0.0016  &0.0014 &\\
     IPC $\alpha$           &\%  &1.6884  &1.6830  &1.6877  &0.0070  &0.0042  & \\
     IPC $\alpha_{\rm H}$&\%  &1.6330  &1.6243  &1.6342  &0.0046  &0.0021 &  \\
     IPC $\alpha_{\rm V}$&\%  &1.7439  &1.7417  &1.7411  &0.0053 &0.0036 &  \\
     IPC $\alpha_{\rm D}$&\%  &0.1881  &0.1835  &0.1846  &0.0029 &0.0009 &  \\
     Non-linearity $\beta_{\rm ramp}$  &ppm/e  &0.5830  &0.6332  &0.6351  &0.0010  &0.0010 &0.0313  \\
     \hline
     \multicolumn{8}{c}{Alternative setups} \\
    Gain $g$               &e/DN &2.0645  &2.0647  &2.0637  &0.0016  &0.0014 & &IQR 85 \\
     IPC $\alpha$           &\% &1.6810  &1.6769  &1.6806  &0.0073  &0.0042 & &$\epsilon=0.025$\\
     Non-linearity $\beta_{\rm ramp}$  &ppm/e &0.5665  &0.6209  &0.6251  &0.0010  &0.0010 & &Not ref. pix. corr.\\
     Non-linearity $\beta_{\rm ramp}$  &ppm/e &0.5816  &0.6321  &0.6342  &0.0010  &0.0010 & &Error mode `none'\\
     Non-linearity $\beta_{\rm ramp}$  &ppm/e &0.5726  &0.6200  &0.6211  &0.0015  &0.0010 & &Error mode `nlipc'\\
    \hline
    \multicolumn{8}{c}{Alternative intervals} \\
    Non-linearity $\beta_{\rm ramp}$  &ppm/e &0.5701  &0.6488  &0.6551  &0.0014  &0.0014 &0.0455 & Frames 3,7,9 \\
    Non-linearity $\beta_{\rm ramp}$  &ppm/e &0.6377  &0.6611  &0.6650  &0.0010  &0.0010 &0.0330 &Frames 3,19,21 \\
    \hline
    \multicolumn{8}{c}{Non-overlapping correlation function (Method 1)} \\
    \multicolumn{8}{c}{BFE+NL-IPC Coefficients - frames 3,11,13,21, baseline-corrected, error mode `bfe'} \\
    $[K^{2}a^{\prime}+KK^{\prime}]_{0,0}$ &ppm/e &-1.2004  &-1.0936  &-1.0772  &0.0154  &0.0045 &0.0541 &Central pixel \\
    $[K^{2}a^{\prime}+KK^{\prime}]_{<1,0>}$ &ppm/e &0.2142  &0.2232  &0.2241  &0.0074  &0.0020 & &Nearest neighbor \\
    $[K^{2}a^{\prime}+KK^{\prime}]_{<1,1>}$ &ppm/e &0.0468  &0.0449  &0.0489  &0.0074  &0.0020 & &Diagonal \\
    $[K^{2}a^{\prime}+KK^{\prime}]_{<2,0>}$ &ppm/e &0.0103  &0.0186  &0.0166  &0.0073  &0.0020 & & \\
    $[K^{2}a^{\prime}+KK^{\prime}]_{<2,1>}$ &ppm/e &0.0018  &0.0102  &0.0042  &0.0052  &0.0015 & & \\
    $[K^{2}a^{\prime}+KK^{\prime}]_{<2,2>}$ &ppm/e &-0.0076  &0.0013  &0.0014  &0.0073  &0.0020 & & \\
    \multicolumn{8}{c}{BFE+NL-IPC Coefficients - frames 3,7,9,13 baseline-corrected, error mode `bfe'} \\
    $[K^{2}a^{\prime}+KK^{\prime}]_{0,0}$ &ppm/e &-1.0446  &-0.9054  &-0.9128  &0.0303  &0.0088 &0.0786 &Central pixel \\
    $[K^{2}a^{\prime}+KK^{\prime}]_{<1,0>}$ &ppm/e &0.2548  &0.2845  &0.2375  &0.0155  &0.0042 & &Nearest neighbor \\
    $[K^{2}a^{\prime}+KK^{\prime}]_{<1,1>}$ &ppm/e &0.0713  &0.0584  &0.0522  &0.0159  &0.0042 & &Diagonal \\
    \multicolumn{8}{c}{BFE+NL-IPC Coefficients - frames 3,19,21,37 baseline-corrected, error mode `bfe'} \\
    $[K^{2}a^{\prime}+KK^{\prime}]_{0,0}$ &ppm/e &-1.2276  &-1.1731  &-1.1637  &0.0077  &0.0049 &0.0572 &Central pixel \\
    $[K^{2}a^{\prime}+KK^{\prime}]_{<1,0>}$ &ppm/e &0.2078  &0.2100  &0.2101  &0.0034  &0.0009 & &Nearest neighbor \\
    $[K^{2}a^{\prime}+KK^{\prime}]_{<1,1>}$ &ppm/e &0.0496  &0.0485  &0.0530  &0.0034  &0.0010 & &Diagonal \\
    \multicolumn{8}{c}{BFE+NL-IPC Coefficients - frames 3,11,13,21, baseline-corrected, error mode `none'} \\
    $[K^{2}a^{\prime}+KK^{\prime}]_{0,0}$ &ppm/e &-1.2086  &-1.1018  &-1.0853  &0.0155  &0.0046 & &Central pixel \\
    $[K^{2}a^{\prime}+KK^{\prime}]_{<1,0>}$ &ppm/e &0.2155  &0.2248  &0.2257  &0.0074  &0.0020 & &Nearest neighbor \\
    $[K^{2}a^{\prime}+KK^{\prime}]_{<1,1>}$ &ppm/e &0.0468  &0.0449  &0.0489  &0.0074  &0.0020 & &Diagonal \\
    \multicolumn{8}{c}{BFE+NL-IPC Coefficients - frames 3,11,13,21, baseline-corrected, error mode `nlipc'} \\
    $[K^{2}a^{\prime}+KK^{\prime}]_{0,0}$ &ppm/e &-1.1827  &-1.0760  &-1.0600  &0.0152  &0.0045 & &Central pixel \\
    $[K^{2}a^{\prime}+KK^{\prime}]_{<1,0>}$ &ppm/e &0.2045  &0.2123  &0.2128  &0.0072  &0.0020 & &Nearest neighbor \\
    $[K^{2}a^{\prime}+KK^{\prime}]_{<1,1>}$ &ppm/e &0.0468  &0.0449  &0.0489  &0.0074  &0.0020 & &Diagonal \\
    \hline
    \multicolumn{8}{c}{Mean-variance relation (Method 2)} \\
    $\hat{a}_{0,0,M2}$ &ppm/e &-1.2843  &-1.0765  &-1.0791  &0.0398  &0.0089 &0.0939 & \\
    $\beta-4(1+3\alpha)\alpha^{\prime}$ &ppm/e &0.3957  &0.3769  &0.3790  &0.0234  &0.0050 & & \\
    $\sum_{a}-8(1+3\alpha)\alpha^{\prime}$ &ppm/e &-0.3747  &-0.5125  &-0.5122  &0.0469  &0.0100 &0.0626 & \\
    \hline
    \multicolumn{8}{c}{Adjacent pixel correlations (Method 3)} \\
    $[K^{2}a^{\prime}+2KK^{\prime}]_{<0,1>}-\alpha\sum_{a}$ &ppm/e &0.2194  &0.2123  &0.2067  &0.0073  &0.0020 & &
    \end{tabular}
    \caption{Averaged results for SCA 18237, based on stacks of flat ramps.  Advanced characterization with \texttt{ncycle=3.}}
    \label{tab:sca18237_full}
\end{table}

\section{Brighter-fatter effect measurements}
\label{sec:bfe}

\subsection{IPNL determination via the non-overlapping correlation function}
\label{ss:char-ipnl}

\changetext{We measure the IPNL parameterized by $[K^{2}a^{\prime}+KK^{\prime}]_{\Delta i, \Delta j}$  via the correlation function for non-overlapping time slices (see red box in Figure~\ref{fig:flow} and Section 5.2 in Paper I for details) over pixel separations $(\Delta i, \Delta j)$ to} a maximum separation of 2 pixels in both horizontal and vertical directions. We provide averages of coefficients related by grid symmetries in Table~\ref{tab:sca18237_full} for a ``fiducial'' \changetext{choice of} frames 3,11,13,21 \changetext{for each of the three groupings of flats: 1st,n3, 2nd,n3, and fid,n23}.

\begin{figure}
    \centering
    \subfloat{{\includegraphics[width=0.47\textwidth]{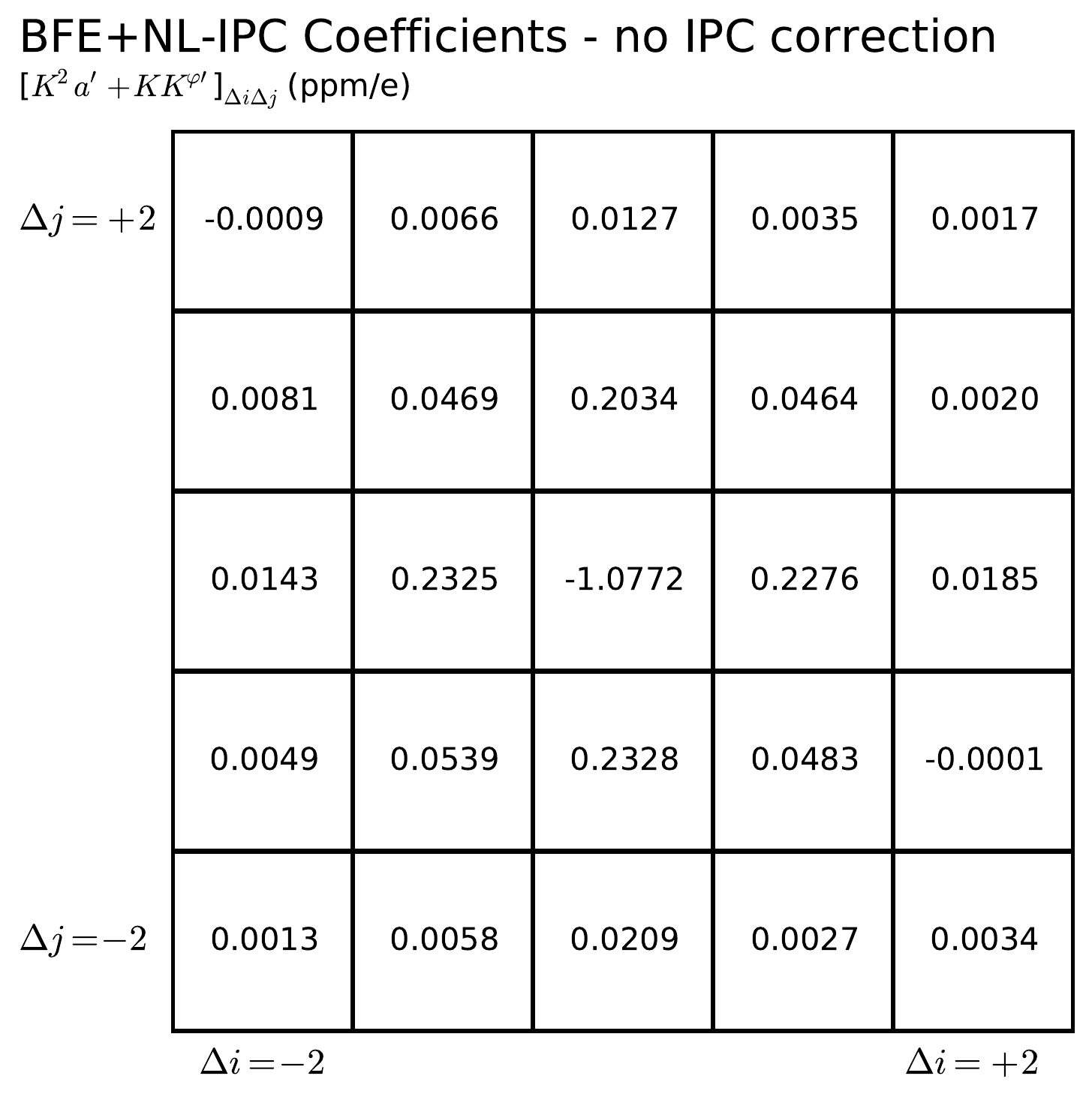}}}%
    \qquad
    \subfloat
    {{\includegraphics[width=0.47\textwidth]{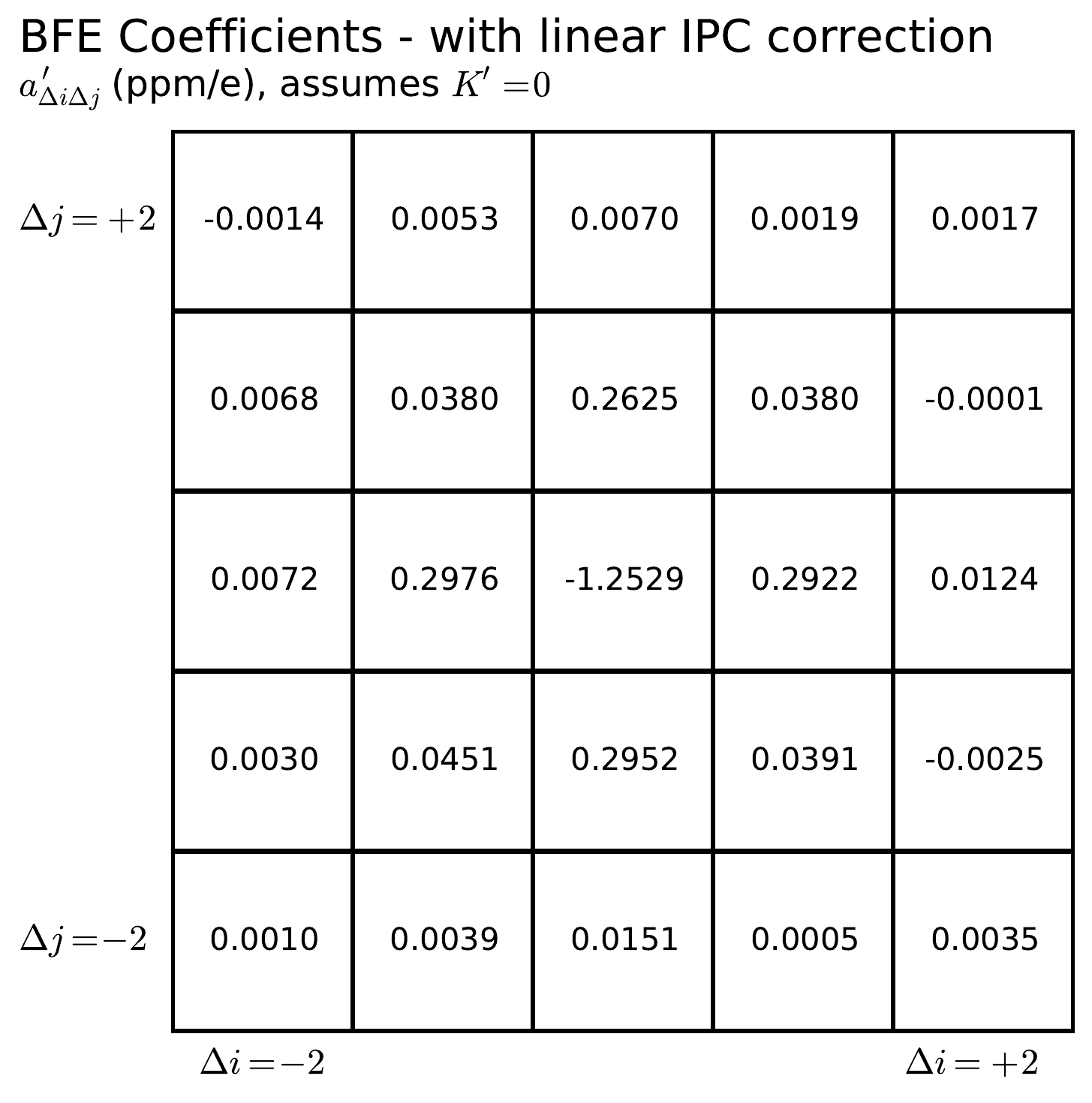}} }%
    \caption{The Method 1 BFE+NL-IPC coefficients (left panel) and IPC-corrected coefficients (right panel) for SCA 18237. These coefficients were measured on a stack of 23 flats, with the full characteristics given in the third column on Table~\ref{tab:sca18237_full}.  Note that the IPC-corrected coefficients assume that the IPC is linear, i.e. the non-overlapping correlations are ascribable entirely to the BFE and not NL-IPC.  The 1$\sigma$ statistical uncertainty for each coefficient is 0.0045 ppm/e (left panel) and 0.0051 ppm/e (right panel).  The central value at zero lag $(\Delta i, \Delta j)=(0,0)$ carries an additional systematic uncertainty of 0.0531 ppm/e (left panel) and 0.0603 ppm/e (right panel) propagated from a standard deviation in $\beta_{\rm ramp}$ of 0.0307 ppm/e.}%
    \label{fig:18237_methone_bfecomp}%
\end{figure}

We also visualize the coefficients for SCA 18237 in the left panel of Figure~\ref{fig:18237_methone_bfecomp}.  The zero-lag coefficient at $(\Delta i, \Delta j)=(0,0)$ has a mean value over all good super-pixels of $-1.077 \pm 0.004 \pm 0.054$ ppm/e.  The first error is the 1$\sigma$ statistical uncertainty, and the second error is obtained from propagating the systematic error in the measurement of $\beta_{\rm ramp}$. The four nearest neighbors with $(\Delta i, \Delta j)$ of $(\pm 1,0)$, and $(0,\pm 1)$ have a mean value of $0.224 \pm 0.002$ ppm/e and the four diagonal neighbors have a mean value of $0.045 \pm 0.002$.  These are high S/N measurements ($>18\sigma$) that give evidence for the existence of inter-pixel non-linearities in this detector, although as this part of the analysis is sensitive to a combination of the BFE and NL-IPC, the exact mechanism cannot yet be determined.  We note that in Paper I, for a simulation based on parameters similar to SCA 18237, we found biases on the extracted BFE coefficients of 12\% in the central component and 2.7\% in the nearest neighbors and determined the cause to be likely related to exclusion of higher order interactions in the current formalism.  The exact contributions of these higher order terms will be revisited in future work.

The right panels of Figure~\ref{fig:18237_methone_bfecomp} explore the scenario in which there is no NL-IPC.  For this case where $K^{\prime}=0$, we compute an order-$\alpha$ inverse kernel and convolve it with $[K^{2}a^{\prime}]$;  the inverse kernel $K^2$ is given by $[K^{-2}]_{0,0}=1+8\alpha$, $[K^{-2}]_{\pm 1,0}=-2\alpha_{\rm H}$, and $[K^{-2}]_{0,\pm 1}=-2\alpha_{\rm V}$.  If the BFE were wholly responsible for the IPNL, the BFE coefficient at zero lag would be given by $-1.253 \pm 0.005 \pm 0.060$ ppm/e.

The BFE kernel for CCDs has been found to be long-range: for example, for DECam, $a_{\Delta i,\Delta j} \propto r^{-\nu}$, where $r=\sqrt{\Delta i^2+\Delta j^2}$ is the pixel separation and $\nu\approx 2.5$ \citep{2015JInst..10C5032G}. For this power law model, the longest possible range is $\nu\rightarrow 2$ (where the total area defect diverges at large $r$), and the shortest possible range is $\nu\rightarrow \infty$ (all missing area from the central pixel appears in the four nearest neighbors). From a fit to the coefficients in the right panel of Figure~\ref{fig:18237_methone_bfecomp}, we find $\nu = 5.6\pm 0.2$ (1$\sigma$ errors based on $\Delta\chi^2$). The BFE in this HgCdTe detector is thus much shorter range than for a CCD.

CCDs have also shown an asymmetry between the ``row'' and ``column'' directions in the BFE. We characterize this quadrupole asymmetry by writing $a_{\rm H} = (a_{1,0}+a_{-1,0})/2$ and $a_{\rm V} = (a_{0,1}+a_{0,-1})/2$. If we interpret the IPNL kernel as BFE, we find
\begin{equation}
a_{\rm H} - a_{\rm V} = 0.016\pm 0.004\,{\rm ppm/e}
~~~{\rm or}~~~
\frac{a_{\rm H} - a_{\rm V}}{a_{\rm H}+a_{\rm V}} = 0.028\pm 0.006.
\end{equation}
The BFE kernel is thus much more symmetrical than has been reported for some CCDs \citep[e.g.][]{2018AJ....155..258C}. There is a $\sim 4\sigma$ detection of an asymmetry; further investigation will be needed to establish whether this small asymmetry is in fact due to the BFE, or due to some other sub-dominant effect.

We compare the IPNL results for the fiducial 3,11,13,21 frames to two other choices of non-overlapping time slices, 3,7,9,13 and 3,19,21,37.  These numbers are also given in Table~\ref{tab:sca18237_full}.  Focusing on the zero-lag coefficient, we have from Eq.~\changetext{(58) from Paper I} that $[K^{2}a^{\prime}+KK^{\prime}]_{0,0}=\frac{g^2}{I^2 t_{ab} t_{cd}}C_{abcd}(0,0) + 2(1-8\alpha)\beta$.  The value of \changetext{$\beta_{\rm ramp}$} is slightly smaller for the shorter time interval, and larger for the longer time interval;  however the difference in $2\changetext{\beta_{\rm ramp}}$ is much smaller than the corresponding differences in $[K^{2}a^{\prime}+KK^{\prime}]_{0,0}$ for the shorter and longer time intervals.  We suggest that the cause for the differences in $[K^{2}a^{\prime}+KK^{\prime}]_{0,0}$ arises from the exclusion of higher order terms in the correlation formalism used in this analysis.

\subsection{The mean-variance relation}
\label{ss:mean-var-relation}
Section 5.5 of Paper I describes two tests that can aid our interpretation of the IPNL detections of \S\ref{sec:bfe}.  In particular, Eq. (\changetext{62}) of Paper I gives two time dependencies for the observed raw gain: one part depends on the start time $t_{a}$ and the classical non-linearity $\beta$, while the other part depends on the duration pattern $t_{ab}$ and $t_{ad}$, $\beta$ and $a_{0,0}$.

\subsubsection{Raw gain vs interval duration}
In the first test, we measure the mean variance slope $\hat g^{\rm raw}_{abad}$ (Eq.~\changetext{62} from Paper I), fix $t_{a}$, and vary $t_{ab}$ and $t_{ad}$.  We fit an intercept $C_0$ and a slope $C_1$ to the equation $\ln \hat g^{\rm raw}_{abad} = C_0 + C_1 I(t_{ad}+t_{ab})$.  Eq.~(\changetext{62}) enables three different interpretations of $C_1$ in the cases of no IPNL, IPNL consisting purely of the BFE, and IPNL consisting purely of NL-IPC:
\begin{equation}
C_1 = 3\beta -(1+8\alpha) [K^2a]_{0,0} + 8(1+3\alpha)\alpha'
= \left\{\begin{array}{lll}
3\changetext{\beta_{\rm ramp}} & & {\tt none} \\
3\changetext{\beta_{\rm ramp}} - (1+8\alpha) [K^2a]_{0,0} & & {\tt bfe} \\
3\changetext{\beta_{\rm ramp}} - 2(1+8\alpha) [KK']_{0,0} & & {\tt nlipc}
\end{array}\right.,
\label{eq:Method-2a}
\end{equation}
Re-arranging the left part of Eq.\ref{eq:Method-2a} and substituting $\changetext{\beta_{\rm ramp}}=\beta-\frac{1}{2}\suma$, we can also define a quantity
\begin{eqnarray}
\hat{a}_{0,0,\rm M2} &\equiv& a_{0,0}+8\alpha a_{<1,0>}-\frac{3}{2}\suma -8(1+3\alpha)\alpha^{\prime} = 3\changetext{\beta_{\rm ramp}}-C_{1} \\ 
&=& \left\{\begin{array}{lll}
a_{0,0}+8\alpha a_{<1,0>} & &{\tt bfe} \\
-8(1+3\alpha)\alpha^{\prime}=2(1+8\alpha)[KK^{\prime}]_{0,0} & &{\tt nlipc}
\end{array}\right.,
\label{eq:Method-2a_a00}
\end{eqnarray}
which is sensitive to the BFE coefficient in the central pixel but also contains a contribution from NL-IPC.

\begin{figure}[ht]%
    \centering
    \subfloat{{\includegraphics[width=0.45\textwidth]{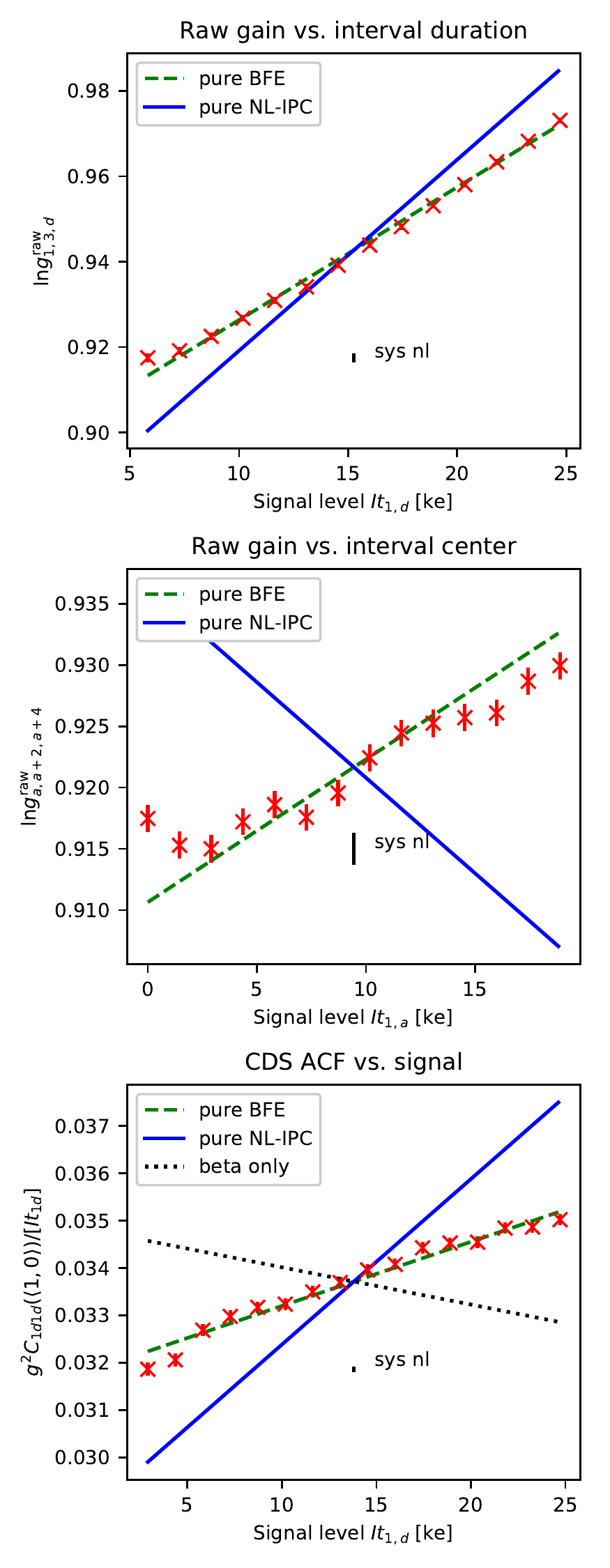}}}%
    \caption{Visual comparison of BFE predictions from Method 1 vs measurements from Methods 2 and 3 for SCA 18237. We only use first flats of a set, with 3 available for SCA 18237.}%
    \label{fig:bfe_m23}%
\end{figure}

The raw gain is computed for frame triplets from [1,3,5], [1,3,6],$...$,[1,5,18], yielding 14 values that are plotted in the top row of Figure~\ref{fig:bfe_m23} for SCA 18237 as a function of the signal level accumulated between the first time slice and the time slice $d=5...18$.  The value of each data point is given by the mean over all super-pixels, with errors on the mean.  The IPNL at zero-lag measured from Method 1 is used to compute the slopes for the pure BFE and pure NL-IPC interpretations, with the central value passing through the center of the measured data points (i.e. the intercept for these slopes is unimportant). A systematic error related to the modeling of the non-linearity (``sys nl'') is also indicated in each panel of Figure~\ref{fig:bfe_m23}. This is based on fitting a 5th order polynomial to the median signal levels in the detector. For both this curve and the quadratic ($\beta$) model, we computed the expected raw logarithmic gain $\ln g^{\rm raw}_{a,b,d}$ for Poisson statistics\footnote{The formula can be derived following the procedure in Paper I for any signal $S(t)$. It is:\\$g^{\rm raw}_{a,b,d} = [S(t_d)-S(t_b)]/\{-2t_aS'(t_a) S'(t_d)[S'(t_d)-S'(t_b)] + t_d[S'(t_d)]^2-t_b[S'(t_b)]^2\}$, where $S$ is the signal curve and $S'$ is its derivative.}, compute the difference, and plot an error bar showing the peak$-$valley range. Note that the absolute gain does not enter because we are using $\ln g^{\rm raw}_{a,b,d}$.  This procedure is intended only to give an indication of the magnitude of systematic errors due to deviation of the classical non-linearity from the $\beta$ model, and in this paper we have not attempted any corrections.

SCA 18237 appears consistent with a pure BFE interpretation within systematic error.  We can quantitatively compare the various estimates for $\hat{a}_{0,0,M2}$ using results from Method 1 as described in Eq.~\ref{eq:Method-2a_a00}.  We have $\hat{a}_{0,0,M2}=-1.2142 \pm 0.0051 \pm 0.0603$ ppm/e (BFE) vs $-2.4453 \pm 0.0102 \pm 0.1228$ ppm/e (NL-IPC) compared with the measured value of $-1.2843 \pm 0.0398 \pm 0.0939$ ppm/e.  Thus, $\hat{a}_{0,0,M2}$ is quantitatively consistent with the pure BFE interpretation.

\subsubsection{Raw gain vs interval center}
In the second test, we measure the mean variance slope, fix $t_{ab}$ and $t_{ad}$, and vary $t_{a}$, fitting an intercept $C'_0$ and slope $C'_1$ to $\ln \hat g^{\rm raw}_{abad} = C'_0 + C'_1 I t_a$.  As for the previous test, we can use the detected IPNL from Method 1 to inform different interpretations of the slope where
\begin{equation}
C'_1 = 2\beta - 8(1+3\alpha)\alpha'
= \left\{\begin{array}{lll}
2\changetext{\beta_{\rm ramp}} & & {\tt none} \\
2\changetext{\beta_{\rm ramp}} & & {\tt bfe} \\
2\changetext{\beta_{\rm ramp}} - 2(1+8\alpha) [KK']_{0,0} & & {\tt nlipc}
\end{array}\right..
\label{eq:Method-2b}
\end{equation}
For this test, $C'_1$ is only sensitive to NL-IPC, as the {\tt none} and {\tt bfe} cases give identical predictions.

We can re-write the above to isolate interesting quantities.  First, we have $\beta - 4(1+3\alpha)\alpha^{\prime} = \frac{1}{2} C_{1}^{\prime}$, which is an alternate way of determining the non-linearity with no leading-order sensitivity to $\suma$ albeit a dependence on NL-IPC.  We can also isolate a combination of $\suma$ and $\alpha^{\prime}$:
\begin{equation}
    \suma - 8(1+3\alpha)\alpha^{\prime} = C_{1}^{\prime}-2\changetext{\beta_{\rm ramp}}.
\end{equation}

The raw gain is computed for frame triplets from [1,3,5], [2,4,6],$...$,[14,16,18], yielding 14 values that are plotted in the middle row of Figure~\ref{fig:bfe_m23} for SCA 18237 as a function of the signal level accumulated between the first time slice and the time slice $a=1,...,14$.  As before, the {\tt none} and {\tt bfe} slopes are plotted.  While the data points seem to prefer the {\tt bfe} slope, there is clearly a change in slope at both low and high signal levels, which warrants further investigation in future studies.

For the 1st flats with SCA 18237, we measure $\beta - 4(1+3\alpha)\alpha^{\prime}=0.3957 \pm 0.0234$, which is inconsistent with the $\changetext{\beta_{\rm ramp}}$ value from basic characterization of $0.583 \pm 0.001 \pm 0.032$ given a hypothesis that the NL-IPC contribution from $\alpha^{\prime}$ is 0.  Then $\suma - 8(1+3\alpha)\alpha^{\prime}=-0.3746 \pm 0.0469 \pm 0.0642$.  As in the previous subsection, if NL-IPC were purely responsible for the measurements we made in Method 1, $8(1+3\alpha)\alpha^{\prime}=-2.7251 \pm 0.0350 \pm 0.1226$ ppm/e, which is inconsistent with the measurement. Taken at face value, this test suggests that NL-IPC is indeed present, with the sign that IPC increases with signal level, but that it is not the dominant mechanism -- it explains $14\pm2\pm2\%$ of the IPNL signal measured in Method 1. However, given that higher-order terms appear to be biasing the Method 1 measurement by $\sim 10\%$ (see Paper I), we urge caution in interpreting this result. We believe the hot pixel test (\S\ref{sec:hotpix}) gives stronger evidence in constraining the NL-IPC.

\subsection{Adjacent pixel correlations}

This method uses the equal-interval correlation function in adjacent pixels as given by Eq. (\changetext{52}) and Eq. (\changetext{55}) for a time-translation-averaged version in Paper I.  As summarized in Sections 3.8.1 and 5.4.3 of Paper I, we can fix the starting time $t_a$ and fit the combination $g^2C_{abab}(\pm 1,0)/(It_{ab})$ as a function of $t_{ab}$, fitting $\frac{g^2}{It_{ab}}C_{abab}(\langle \pm 1,0 \rangle) = C''_{0} + C''_{1} It_{ab}$.  The slope is given by
\begin{equation}
C''_{1} =  -8\alpha\beta + \alpha \suma + [K^2a]_{\langle 1,0\rangle} +2 [KK']_{\langle 1,0\rangle}
= \left\{\begin{array}{lll}
-8\alpha\changetext{\beta_{\rm ramp}} & & {\tt none} \\
-8\alpha\changetext{\beta_{\rm ramp}} + [K^2a']_{\langle 1,0\rangle} & & {\tt bfe} \\
-8\alpha\changetext{\beta_{\rm ramp}} + 2[KK']_{\langle 1,0\rangle} & & {\tt nlipc} \end{array}\right..
\label{eq:Method-3}
\end{equation}
We can add $8\alpha_{\rm H}\changetext{\beta_{\rm ramp}}$ to the left hand part of Eq.~\ref{eq:Method-3} to obtain
\begin{equation}
    C''_{1} +8\alpha\changetext{\beta_{\rm ramp}} = [K^2a]_{\langle 1,0\rangle} +2[KK']_{\langle 1,0\rangle} - 3\alpha \suma = [K^2a'+2KK']_{\langle 1,0\rangle}-\alpha \suma
\end{equation}

IPC is measured via basic characterization of frame triplets from [1,2,3], [1,2,4],..., [1,2,18], and CDS auto-correlations are computed for [frame 3 - frame 1], [frame 4 - frame 1],..., [frame 18 - frame 1].
The bottom panel of Figure~\ref{fig:bfe_m23} shows the measurements from Method 3 with predictions from Method 1 over plotted. SCA 18237 agrees well with the pure BFE interpretation.

We can compare the estimate of $[K^2a'+2KK']_{\langle 1,0\rangle}-\alpha \suma$ with the Method 1 result of $[K^2a'+KK']_{\langle 1,0\rangle}$.  For a pure BFE interpretation, these quantities would be equivalent;  for a pure NL-IPC interpretation, the two would differ by a factor of 2.  For SCA 18237, this test measures $0.2067 \pm 0.0020$ ppm/e vs the Method 1 value of $0.2241 \pm 0.0020$ ppm/e. This gives a ratio of $0.922\pm0.012$, which should be 1 for pure BFE and 2 for pure NL-IPC. This favors the BFE interpretation, although the difference from 1 is statistically significant with 23 flats.

\subsection{Comparison to IPC measured on hot pixels}
\label{sec:hotpix}

An alternative method to assess the IPC is to use hot pixels observed during dark exposures. The method relies on the fact that if the pixel $(i,j)$ is hot (i.e., the photodiode leaks significant current even in the absence of illumination), then a signal (in DN) will appear in the neighboring pixels due to capacitive coupling. This method is in principle more direct than the flat field method; it does not involve the BFE or other sources of correlations between pixels. It also enables one to explore a wide range of signal levels, including very low signal levels where control of systematics is difficult with flats. The main drawback is that it only probes the specific pixels that are hot, and one must beware of issues involving hot pixel selection \changetext{and the possibility that hot pixels may behave differently from the science-grade pixels in ways other than being hot}. \changetext{The procedure described here is qualitatively similar to the one used in \citet{2011wfc..rept...10H} to make on-orbit measurements of the IPC for the infrared channel of the Wide Field Camera 3.}

The {\sc solid-waffle} system selects hot pixels as follows. First, for each dark D$_k$, we make the CDS image $S_{1,65}(i,j|{\rm D}_k)$ from the 1st and 65th time frames. We provisionally select pixels to be ``hot" if the average $M_{1,65}(i,j) = \frac1{N_{\rm dark}}\sum_{k=1}^{N_{\rm dark}}  S_{1,65}(i,j|{\rm D}_k)$ of these images is in a given signal range (specified as a minimum and maximum DN, e.g., 1000--2500). We next impose additional cuts on the pixels to ensure that they are {\em isolated} (so that our IPC measurements are not affected by other nearby hot pixels) and {\em repeatable} (so that we are not selecting cosmic rays or pixels affected by random telegraph noise). We impose the isolation cut first since we also want to be isolated from unstable pixels. This cut requires the $5\times 5$ block centered on it to have no other pixel that is brighter than $\epsilon_{\rm i}$ times the pixel itself, i.e.,
\begin{equation}
M_{1,65}(i + \Delta i, j+\Delta j) < \epsilon_{\rm i}M_{1,65}(i,j) ~~~{\rm for}~~~ |\Delta i|\le 2,~|\Delta j|\le 2, ~(\Delta i,\Delta j)\neq (0,0).
\end{equation}
The default is $\epsilon_{\rm i} = 0.1$. We also require this $5\times 5$ block to contain no reference pixels (i.e., to not be in the first or last 6 rows or columns).
The second cut requires that the candidate hot pixel be repeatable in the $N_{\rm dark}$ darks:
\begin{equation}
\max_{k=1}^{N_{\rm dark}} S_{1,b}(i,j|{\rm D}_k) - \min_{k=1}^{N_{\rm dark}} S_{1,b}(i,j|{\rm D}_k)
\le \epsilon_{\rm r} \frac1{N_{\rm dark}}\sum_{k=1}^{N_{\rm dark}} S_{1,65}(i,j|{\rm D}_k)
~~~ {\rm for} ~~2\le b\le 65,
\end{equation}
where $\epsilon_{\rm r}$ is a repeatability parameter (default: 0.1). Note that we impose this criterion for intermediate frames $b$: we want pixels that exhibit the same time history in every dark exposure.

A hot pixel can be used to give an estimate of the IPC using CDS images from any final frame $b$ -- that is, from $S_{1,b}(i,j|{\rm D}_k)$. This is useful because by varying $b$, we may determine how the IPC varies with signal level. The steps are as follows:
\begin{list}{$\bullet$}{}
\item $[${\em Optional, default = on}$]$: Perform a lowest-order non-linearity correction on the CDS frames, $S_{1,b}(i,j|{\rm D}_k) \rightarrow S_{1,b}(i,j|{\rm D}_k) [1 + \beta g S_{1,b}(i,j|{\rm D}_k)]$, where $\beta g$ is the product of non-linearity and gain (units: DN$^{-1}$).
\item We construct the median of the dark frames: $M_{1,b}(i,j)$.
\item If the pixel $(i,j)$ is hot, then we extract the $3\times 3$ postage stamp $M_{1,b}(i+\Delta i,j+\Delta j)$ for $|\Delta i|, |\Delta j|\le 1$. We also extract a ``background'' estimate $B$ from the mean of the surrounding $5\times 5-3\times 3 =16$ pixels.
\item We construct the IPC kernel for that hot pixel from:
\begin{equation}
\hat K(\Delta i,\Delta j) = \frac{M_{1,b}(i+\Delta i,j+\Delta j) - B}{\sum_{\Delta i'=-1}^1 \sum_{\Delta j'=-1}^1 [
M_{1,b}(i+\Delta i',j+\Delta j') - B ]}.
\end{equation}
Averages of 4 pixels such as $\alpha$ (average of $K$ in 4 nearest neighbors) and $\alpha_{\rm D}$ (4 diagonal neighbors) can also be reported.
\end{list}
The aforementioned procedure gives an estimate of $\alpha$ for each selected hot pixel and each time sample $b$.

\begin{figure}
    \centering
    \includegraphics[width=6in]{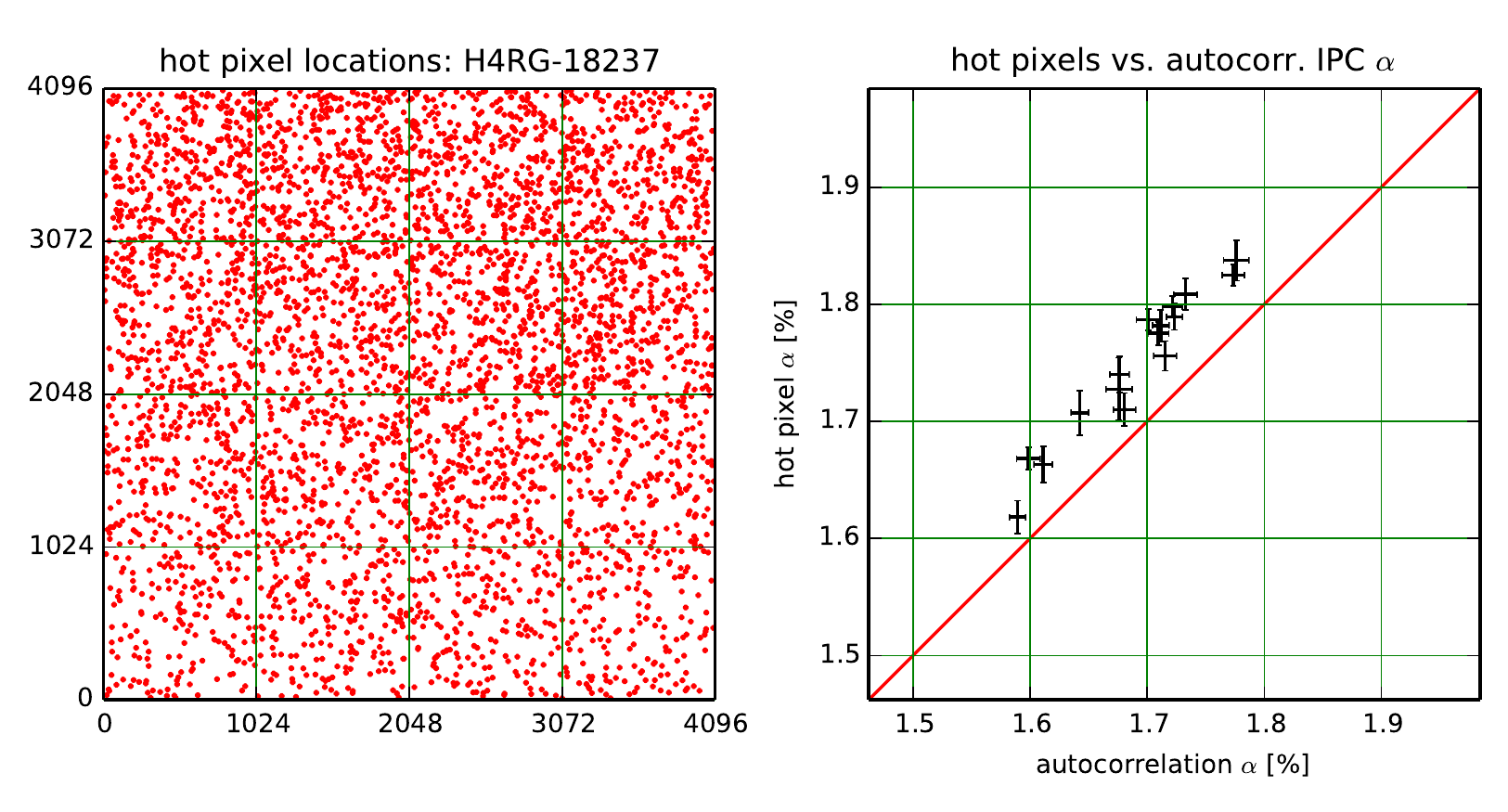}
    \caption{\label{fig:hotpix}Hot pixel IPC analysis for SCA 18237 (5 dark exposures, 6604 hot pixels). Parameters are: hot pixel brightness 1000--2000 DN; $\epsilon_{\rm i}=0.1$; and $\epsilon_{\rm r} = 0.1$. {\em Left panel}: the locations of selected hot pixels. {\em Right panel:} The comparison of median IPC from hot pixels (vertical axis) versus the advanced auto-correlation analysis (horizontal axis), binned into $1024\times 1024$ sub-regions on the SCA (so that there are 16 data points). The hot pixel error bars are from the binomial distribution, whereas the auto-correlation errors are $1\sigma$ errors on the mean.}
\end{figure}

In Figure~\ref{fig:hotpix}, we show plots from {\sc solid-waffle} for SCA 18237. The results here are for hot pixels in the 1000--2000 DN range. We used only the first 5 dark frames for this analysis. We divided the SCA into 16 $1024\times 1024$ sub-regions and compared the hot pixel to auto-correlation IPC measurements in each sub-region. One can see that the results are tightly correlated: the two methods are measuring the same spatial structure, which is reassuring given that they used completely different stimuli to measure the cross-talk (dark current versus flat illumination). One can also see that there is an offset between the two methods, at the level of $\alpha_{\rm hot\,pix} - \alpha_{\rm autocorr} = 0.06$\%. This could be interpreted as either a systematic error in one or both of the measurements, or a symptom of IPC non-linearity. To investigate the latter, we considered the signal dependence of the hot pixel IPC measurement.

\begin{figure}
    \centering
    \includegraphics[width=6in]{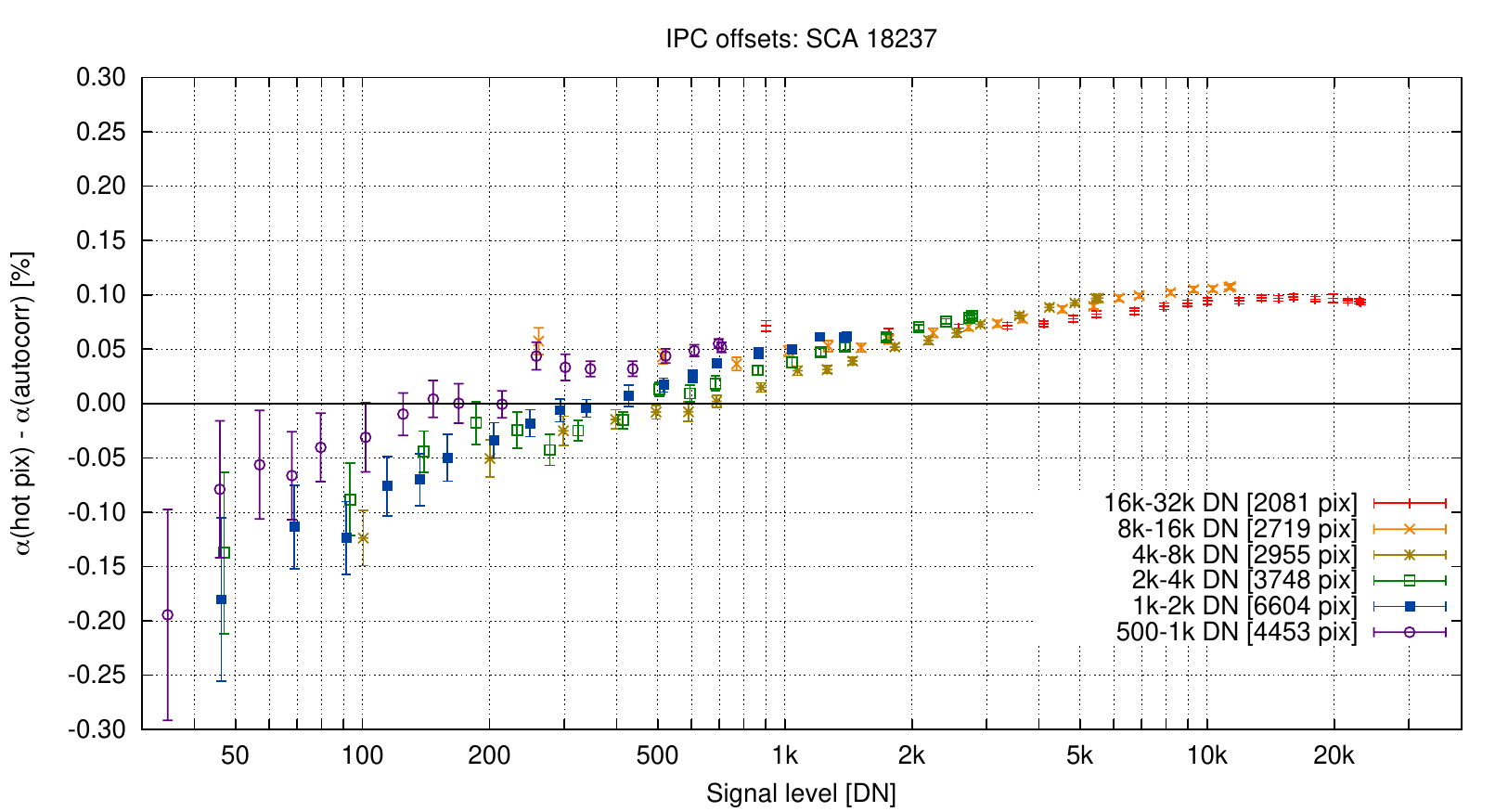}
    \caption{\label{fig:nlipc}The hot pixel IPC as a function of signal level, for SCA 18237. We subtracted the auto-correlation $\alpha$ from the vertical axis, however exactly the same auto-correlation $\alpha$ map was used as a reference for every point (i.e., by construction it has no time or signal dependence). Each point style reflects a selection of pixels, from $S_{1,65} = 16000-32000$ DN (red) through 500--1000 DN (purple). The sequence of points indicates hot pixel measurements from different time samples $S_{1,b}$ (varying $b$; $b=2$ is the first point shown, $b=65$ is the last, and in between we have shown only some of the points -- $b = 1,3,5,7\times 2^n$ for integer $n$ -- to avoid clutter). Errors shown are binomial errors on the median.}
\end{figure}

The signal dependence of the IPC can be explored by both selecting different pixels (from slightly warm to very hot) and by comparing different time stamps. We do both tests in Figure~\ref{fig:nlipc}, but note that the time stamp test has the advantage of not depending at all on the spatial variation of IPC and depending only on the non-linearity. One can see that SCA 18237 shows an increase in $\alpha$ as a function of signal level, which can be seen all the way from tens of DN (comparable to the reset noise) all the way up to $2.4\times 10^4$ DN (roughly half full well).

If NL-IPC were responsible for the entirety of the inter-pixel non-linearity signal observed in Method 1, then we would have to have $[KK']_{\langle 1,0 \rangle} = 0.22$ ppm/e. Expanding the kernel, we see that $[KK']_{\langle 1,0 \rangle} = (1-8\alpha)\alpha'$, so this suggests that the signal dependence of $\alpha$ in DN units would be
\begin{equation}
g\alpha' = \frac{g}{1-8\alpha}[KK']_{\langle 1,0 \rangle}
= 
5.4\times 10^{-7}\,{\rm DN}^{-1}.
\end{equation}
The relevant range of signal levels is up to 15 kDN (the signal level at frame 21), so we would expect $\alpha$ (as inferred from the hot pixels) to change by $5.4\times 10^{-7}\,{\rm DN}^{-1} \times 15\,{\rm kDN} = 0.81\%$ from low values up to 15 kDN. Instead, the differences in Figure~\ref{fig:nlipc} over this range are 0.1--0.2\%. A detailed quantitative comparison is not possible since NL-IPC can depend (in principle) on both signal level and contrast: the flat field measures the dependence on signal level at low contrast, whereas the hot pixel test measures the dependence on signal level in one pixel with a background of near zero. The two measurements need not be exactly the same (and in the formalism of \citealt{2018PASP..130g4503D}, they are not). Nevertheless, the very low NL-IPC that we observe in the hot pixel test suggests that Method 1 is seeing primarily BFE rather than NL-IPC.

\section{Discussion and Future Work}
\label{sec:discussion}

We have analyzed the flat field statistics of a prototype WFIRST H4RG-10 detector array (SCA 18237) following the procedures introduced in Paper I.  In summary, we started with a basic characterization of the detector, where we constructed CDS images per flat per super pixel, computed a median over the flats per super pixel and a mask, performed a reference pixel subtraction, computed the raw gain, horizontal and vertical correlations, and an estimate of ramp curvature.  We then solved for the IPC and non-linearity corrected gain, horizontal and vertical $\alpha$ components, current, and classical non-linearity $\beta$.  The following step was to measure the non-overlapping correlation function, which is almost a direct test for the presence of inter-pixel non-linearities.  We then used the non-overlapping correlations, interpreted as either BFE or NL-IPC, to de-bias the original ``basic characterization'' parameters of $g$, $\alpha$, etc. with an iterative process.  We performed a range of robustness checks to ensure the stability of our results when various analysis choices were modified.  We conducted four complementary investigations to help build the interpretation of the mechanism(s) behind the inter-pixel non-linearities: (1) raw gain vs interval duration, (2) raw gain vs interval center, (3) equal-interval correlation function in adjacent pixels, and (4) IPC measurement on hot pixels.  The main results can be recapitulated as follows.

\begin{itemize}
    \item There is large-scale spatial variation of the IPC at the $\sim 0.3\%$ level. The same spatial variation is observed in both the autocorrelation and hot pixel tests for IPC. There is a $\sim 0.06\%$ overall offset between the two methods that is under investigation. IPC and its spatial variation will be further investigated with single pixel reset tests during WFIRST flight detector acceptance testing. 
    \item We have used the formalism built up in Paper I to detect residual correlations between difference frames of flat fields where the time intervals do not overlap.  SCA 18237 shows a decrement in the central kernel value at high S/N.  While this non-overlapping correlation method provides the highest S/N measurements, the underlying mechanism includes contributions from both the BFE and NL-IPC. If interpreted as pure BFE, this measurement would indicate an Antilogus coupling coefficient to the 4 nearest neighbors of $a_{\langle 1,0\rangle} = 0.287\pm0.003$ ppm/e and to the 4 diagonal neighbors of $a_{\langle 1,1\rangle} = 0.040\pm0.003$ ppm/e.  This effect is of the same order of magnitude compared to \citet{2018PASP..130f5004P}, who use spot illumination on an H2RG device (with different geometry: 18$\mu$m pitch pixels). The differences between their analysis and ours complicate a more quantitative comparison. 
    \item The main effect of the BFE on weak lensing analyses is generally through the stars used to estimate the PSF. The WFIRST Science Requirements Document uses a reference star with a total fluence of $8.7\times 10^4$ collected electrons per exposure. If we use an obstructed Airy disk centered on a pixel center and with no extra spreading due to aberrations, charge diffusion, or image motion (the most extreme case), and the BFE kernel for SCA 18237, then averaged over the duration of the exposure the area of the central pixel is modified by $-2.2\%$ in $J$-band and $-1.5\%$ in $H$-band. Since WFIRST aims for PSF size calibration at the $7.2\times 10^{-4}$ level, the BFE will have to be accurately measured and corrected. \changetext{The WFIRST Wide Field Instrument Calibration Plan presented at the WFIRST System Requirements Review in 2018 estimated that if 5\% (in an RSS sense) of the PSF size error budget is allocated to BFE, then BFE coefficient $a_{<1,0>}$ needs to be measured to an accuracy of $\sigma_{a,<1,0>}({\rm req't}) = 0.0031\,{\rm ppm/e}$.  Here we have achieved a {\em statistical} error on $a_{<1,0>}$ of 0.003 ppm/e.  This is encouraging but is by no means the end of the story since we measured a single IR detector and used low contrast data (fluctuations around a flat); further laboratory studies will be needed to validate the accuracy of the BFE model for the high contrasts expected when observing PSF stars.}
    \item In order to determine whether the BFE or NL-IPC is the dominant mechanism behind the measured IPNL,  we can run other tests such as measuring the dependence of the raw gain on either the start time or the interval duration.  A third test involves the scaling of the adjacent-pixel covariance as a function of signal level. All of these tests favor the BFE rather than NL-IPC as the dominant mechanism.
    \item The hot pixel analysis is a more direct way of assessing the IPC and can be used to investigate a wide range of signal levels. We find evidence for NL-IPC, with the IPC coefficient increasing for greater signal in the hot pixel. However this NL-IPC is at least a factor of $\sim 5$ too small to explain the non-overlapping correlation function measurement. This is consistent with the interpretation that BFE is dominant over NL-IPC.  We note that while we do not find the same level of signal dependence of IPC in this hot pixel analysis as \citet{2016SPIE.9915E..2ID}, we are investigating a detector with different pixel geometry  -- for example, the H4RG-10 has 10$\mu$m pitch pixels compared to 18$\mu$m pitch pixels for the H2RG devices analyzed in \citet{2016SPIE.9915E..2ID}. \changetext{In future work, we will compare the hot pixel results to those obtained from measurements of single pixel reset data.}
    \item The most significant limitation of the present BFE measurement is that the current model for the correlation function $ C_{abcd}(\Delta i,\Delta j)$ keeps only the leading-order non-linear terms, i.e., we use the quadratic $\beta$-model for non-linearity, and drop terms of order $a^2$ or $a\beta$. Simulations show that this induces a bias of $\sim 12\%$ for SCA 18237-like parameters. This limitation is not fundamental, and will be remedied in a future paper.
    \item The BFE kernel is shorter range than observed in thick CCDs, with an observed fall-off of $\propto r^{-5.6\pm 0.2}$. This makes physical sense given the thin geometry of the HgCdTe detectors and agrees qualitatively with the model described in \citet{2017JInst..12C4009P,2018PASP..130f5004P} wherein higher-signal pixels have smaller depletion regions, which is more of a local effect compared to the case of CCDs. There is also only a small horizontal vs.\ vertical asymmetry in the non-overlapping correlation function; if ascribed to the BFE, this asymmetry suggests $(a_{\rm H}-a_{\rm V})/(a_{\rm H}+a_{\rm V}) = 0.028\pm 0.006$.
\end{itemize}

This study is one of the early steps in the long-term effort to calibrate WFIRST detectors.
In the near term, we plan to improve the modeling to include higher-order non-linear terms and a more general classical non-linearity model. We also plan more investigation into the spatial structure of the BFE and IPC, and further characterization of the changes in flat field as the array is exposed to light (the 1st vs.\ 2nd flat effect). We plan to investigate the BFE using \changetext{data with higher contrast than the flat fields analyzed here such as} the focused spot and speckle fringe illumination data that will be acquired on some SCAs during detector characterization testing. \changetext{These are critical because the stars used for PSF determination have high contrast and the BFE model needs to be validated against data in the high-contrast regime.} Since detector characteristics can vary substantially from one device to another (even when built according to the same recipe), we plan to run the {\sc solid-waffle} tools on a larger sample of SCAs, including on flat fields of all of the WFIRST flight candidate SCAs.  The lessons learned will feed back into calibration planning for the WFIRST mission.

\section*{Acknowledgements}

We thank the Detector Characterization Laboratory personnel, Yiting Wen, Bob Hill, and Bernie Rauscher at NASA Goddard Space Flight Center for their efforts enabling the existence and access to the data analyzed in this series of papers, and we thank Chaz Shapiro, Andr\'es Plazas, and Eric Huff for helpful discussions. We thank Jay Anderson and Arielle Bertrou-Cantou for useful presentations to the Detector Working Group on their analyses of non-linearities in the HST/WFC3-IR and Euclid H2RG detectors.   We thank the anonymous referee for helpful suggestions that improved the clarity of this paper. We are also grateful for the use of \citet{OhioSupercomputerCenter1987} for computing the results in this work. AC and CMH acknowledge support from NASA grant 15-WFIRST15-0008. During the preparation of this work, CMH has also been supported by the Simons Foundation and the US Department of Energy.  Software: Astropy \citep{2013A&A...558A..33A,2018AJ....156..123A}, fitsio \citep{fitsio}, Matplotlib \citep{Hunter:2007}, NumPy \citep{numpy}, SciPy \citep{scipy}

\bibliographystyle{aasjournal}
\bibliography{main}

\end{document}